\documentclass[conference]{IEEEtran}
\usepackage{graphicx}
\usepackage{algorithm}
\usepackage{algorithmic}
\usepackage{amsmath}
\newcommand{\pluseq}{\mathrel{+}=}

\usepackage{multicol}
\usepackage{color,soul}

\usepackage{tikz}
\newcommand*\circled[1]{\tikz[baseline=(char.base)]{
\node[shape=circle,draw,inner sep=1pt, color=black, fill=gray] (char) {#1};}}

\newcommand{\squeezeup}{\vspace{-6mm}}
\begin{document}

%\title{A Learning Automata-based Graph Partitioning Algorithm for Cloud Workloads}

\title{Partitioning Graphs for the Cloud using Reinforcement Learning}


\author {
    \IEEEauthorblockN 
    {
        Mohammad Hasanzadeh Mofrad, Rami Melhem
    }
    \IEEEauthorblockA
    {
        Department of Computer Science, University of Pittsburgh \\
        Pittsburgh, USA \\
        Email: \tt\small \{moh18, melhem\}@pitt.edu
    }
    
    \and
    \IEEEauthorblockN {
        Mohammad Hammoud
    }
    \IEEEauthorblockA 
    {
        Carnegie Mellon University in Qatar \\
        Doha, Qatar\\
        Email: \tt\small mhhammoud@cmu.edu
    }
}

\maketitle
\begin{abstract}
    In this paper, we propose Revolver, a parallel graph partitioning algorithm capable of partitioning large-scale graphs on a single shared-memory machine. Revolver employs an asynchronous processing framework, which leverages reinforcement learning and label propagation to adaptively partition a graph. In addition, it adopts a vertex-centric view of the graph where each vertex is assigned an autonomous agent responsible for selecting a suitable partition for it, distributing thereby the computation across all vertices. The intuition behind using a vertex-centric view is that it naturally fits the graph partitioning problem, which entails that a graph can be partitioned using local information provided by each vertex's neighborhood. We fully implemented and comprehensively tested Revolver using nine real-world graphs. Our results show that Revolver is scalable and can outperform three popular and state-of-the-art graph partitioners via producing comparable localized partitions, yet without sacrificing the load balance across partitions.
s    
\end{abstract}


% \IEEEpeerreviewmaketitle

\section{Introduction}\label{intro}

    Graphs are remarkably capable of capturing complex data dependencies present in various real-world domains, including biological networks, web graphs, social networks, and transportation routes, to mention just a few. Extracting useful information and gleaning insights from such graphs (a process denoted as {\em graph analytics}) is becoming central to our modern life. For instance, Facebook continually mines gigantic social networks to determine shared connections, detect communities, and propagate advertisements for massive number of users. Other common graph analytics applications include scene reconstruction from image collections and topical news recommendations from micro-blogging services (e.g., Twitter). 
    %, among others.
    
    %For instance social networks and web graphs can be represented 
    
    %\hl{Social networks are ceaselessly producing big graphs that do not fit into the memory of a single machine. Cloud computing enables processing such big graphs by hosting distributed graph analytics. In distributed graph analytics, the computation load and communication volume of processing of this bulky graphs can be balanced across the Cloud by utilizing a partitioning algorithm. Thus, a graph partitioning algorithm in the Cloud basically partitions or overpartitions a big graph and allocates one or more partitions to each server of the Cloud.}

    The scale of the graphs generated by such applications is significantly increasing, yielding graphs with millions of vertices and edges referred to as \textit{big graphs}, which cannot be typically mined efficiently on a single machine~\cite{ahmad2018la3}. Consequently, for  efficient execution of such applications, large-scale clusters are usually needed~\cite{redekopp2011performance, gonzalez2012powergraph}. Cloud  computing  services like Amazon  EC2~\cite{AmazonEC2}, Microsoft  Azure~\cite{MicrosoftAzure}, and  Google AppEngine~\cite{GoogleAppEngine} offer unprecedented levels of on-demand access to computing and storage resources, allowing thereby large-scale  graph analytics to be effectively pursued.
    
    In addition to resource requirements, efficient graph analytics necessitates platforms tailored specifically for big graphs. The current \textit{graph analytics platforms} can be examined from different angles. More precisely, in terms of computing needs they can be divided into three types: 1) \textbf{single node non-scalable solutions} such as \textit{Gunrock} \cite{wang2016gunrock}, \textit{Ligra} \cite{shun2013ligra}, \textit{Polymer} \cite{zhang2015numa} and \textit{Galois} \cite{nguyen2013lightweight}, 2) \textbf{single node out-of-core solutions} such as \textit{GraphChi} \cite{kyrola2012graphchi}, \textit{X-Stream} \cite{roy2013x}, \textit{GridGraph} \cite{zhu2015gridgraph} and \textit{Mosaic} \cite{maass2017mosaic}, and 3) \textbf{distributed solutions} such as \textit{LA3}~\cite{ahmad2018la3}, \textit{Apache Giraph} \cite{ching2015one}, \textit{Google Pregel} \cite{malewicz2010pregel}, \textit{GraphLab} \cite{low2012distributed}, \textit{PowerGraph} \cite{gonzalez2012powergraph}, \textit{PowerLygra} \cite{chen2015powerlyra} and \textit{PowerSwitch} \cite{xie2015sync}.
    
    %Efficient graph analytics requires platforms tailored for exascale graphs. The current \textit{graph analytics platforms} can be examined from different dimensions. Specifically, in terms of computing capacity they can be divided into three types: 1) \textbf{single node solutions} such as \textit{Gunrock} \cite{wang2016gunrock}, \textit{Ligra} \cite{shun2013ligra}, \textit{Polymer} \cite{zhang2015numa} and \textit{Galois} \cite{nguyen2013lightweight}, 2) \textbf{single node out-of-core solutions} such as \textit{GraphChi} \cite{kyrola2012graphchi}, \textit{X-Stream} \cite{roy2013x}, \textit{GridGraph} \cite{zhu2015gridgraph} and \textit{Mosaic} \cite{maass2017mosaic}, and 3) \textbf{distributed solutions} such as \textit{Apache Giraph} \cite{ching2015one}, \textit{Google Pregel} \cite{malewicz2010pregel}, \textit{GraphLab} \cite{low2012distributed}, \textit{PowerGraph} \cite{gonzalez2012powergraph}, \textit{PowerLygra} \cite{chen2015powerlyra} and \textit{PowerSwitch} \cite{xie2015sync}.
    
    Graph analytics platforms can also be categorized based on programming models, namely: 1) \textbf{vertex-centric} in Pregel \cite{malewicz2010pregel} and Giraph \cite{ching2015one}, 2) \textbf{edge-centric} in \textit{X-stream} \cite{roy2013x}, 3) \textbf{sub-graph-centric} in \textit{GoFFish} \cite{simmhan2014goffish}, and 4) \textbf{graph-centric} in \textit{Giraph++} \cite{tian2013think}. In addition, they can be classified in terms of two major execution models, namely: 1) \textbf{synchronous model}, where vertices progress in a lock-step fashion such as in Giraph \cite{ching2015one}, and 2) \textbf{asynchronous model}, where vertices can change values anytime and be several steps apart during execution such as in PowerGraph \cite{gonzalez2012powergraph}.
    
    Distributed graph analytics platforms, irrespective of their programming and execution models, necessitate partitioning input graphs. A popular technique to partition graphs is referred to as \textit{balanced graph partitioning}, which tries to divide any input graph into a set of roughly equal sub-graphs (or partitions), while reducing edge cuts among pairs of partitions so as to minimize overall communication cost. Clearly, balanced graph partitioning is theoretically NP-hard~\cite{hammoud2014distributed}; however, it can be solved effectively using a vertex-centric approach. To exemplify, \textit{Ja-Be-Ja} \cite{rahimian2013ja}, \textit{Fennel} \cite{tsourakakis2014fennel}, and \textit{Spinner} \cite{martella2017spinner} suggest vertex-centric algorithms for balanced graph partitioning. 
    
    In this paper, we propose {\em \textbf{Revolver}}, an asynchronous single-node graph partitioning algorithm, which adopts a vertex-centric view of graphs. The vertex-centric approach enables Revolver to partition any graph in a parallel fashion. Alongside, the asynchronous execution model allows it to involve the most recent partitioning results into the ongoing computations and, subsequently, produce load-balanced partitions. Furthermore, asynchrony permits it to skip the strict barrier requirements of the synchronous execution model (e.g., the famous BSP framework~\cite{valiant1990bridging}), thus converging quickly.
    
    Within its core, Revolver utilizes Label Propagation (LP) \cite{barber2009detecting} to train Learning Automata (LA) \cite{narendra2012learning} for graph partitioning. LA is a subclass of Reinforcement Learning (RL), which focuses on training autonomous agents in an interactive environment in order to optimize cumulative rewards. LA have applications in \textit{evolutionary optimization} \cite{hasanzadeh2013adaptive}, \textit{Cloud and Grid computing} \cite{hasanzadeh2014grid, hasanzadeh2015distributed, jobava2018achieving}, \textit{social networks} \cite{rezvanian2017new}, \textit{image processing} \cite{mofrad2015cellular}, and \textit{data clustering} \cite{hasanzadeh2017learning}, to mention a few.
    
    In addition to using LA, Revolver introduces: 1) a new highly accurate normalized LP, 2) a weighted learning automaton, which particularly suits graph workloads, and 3) highly balanced partitions compared to existing graph partitioners.
    
    The paper is organized as follows. We formally define the graph partitioning problem in Section \ref{Problem}. A background on LP and LA algorithms is provided in Section \ref{Background}. In Section \ref{revolver}, our weighted learning automaton and partitioning algorithm is presented. The evaluation methodology and results are presented in Section \ref{experimens}. Finally, we conclude in Section \ref{conclusions}.

\section{Problem Definition} \label{Problem}
    To begin with, let us assume a directed graph $G = (V, E)$ where $V$ is the set of vertices and $E$ is the set of outgoing edges (directed edges). An edge-centric $k$-way graph partitioning algorithm divides $E$ into $k$ distinct partitions of almost equal size $|E| / k$ partitions. Let $L = \{B(l) \ | \ l \in \{1, ..., k\}, \ \bigcup_{l=1}^{k} \ B(l) = V \ \land \ \bigcap_{l=1}^{k} B(l) = \emptyset \}$. Consequently, a balanced partitioning assignment can be defined as:
    %\abovedisplayshortskip=-30pt
    %\belowdisplayshortskip=-30pt
    %\abovedisplayskip=-3pt
    %\belowdisplayskip=-3pt
    %\abovedisplayskip=0pt
    %\setlength\abovedisplayskip{-3pt}
    \begin{equation}
         \label{eq:problem}
            \frac{|E|}{k} \bigg(1 - \Big((1 + \epsilon) \cdot (k - 1) \Big) \bigg) \leq \ |b(l)| \ \leq \frac{|E|}{k} \cdot \bigg(1 + \epsilon \bigg)
    \end{equation}
    
    where $b(l)$ is the set of outgoing edges that belongs to $B(l) \subset V$ with $B(l)$ being the subset of vertices assigned to the $l^{th}$ partition and $\epsilon > 0$ is the imbalanced ratio. Also, to guarantee having non-empty partitions, $\epsilon$ should satisfy the following inequality:
    %\abovedisplayshortskip=-5pt
    %\belowdisplayshortskip=-5pt
    %\setlength\abovedisplayskip{3pt}
    \begin{equation}
         \label{eq:epsilonInequality}
         (k - 1) \cdot \epsilon \cdot \frac{|E|}{k} << \frac{|E|}{k} \
         \rightarrow
         (k - 1) \cdot \epsilon << 1
    \end{equation}
    
   Upon running a graph application (e.g. PageRank) in a distributed fashion, the biggest partition bounds the amount of computation and the number of inter-partition edges bounds the amount of communication under each processing step. Hence, a balanced partitioning of a graph workload potentially lowers the runtime of any distributed graph analytics platform via imposing near-uniform utilization across machines while reducing communication. Examples of balanced partitioners are \textit{Kernighan-Lin} \cite{kernighan1970efficient}, \textit{Spectral partitioning} \cite{chan1994spectral}, and \textit{Metis} \cite{karypis1998fast}. Besides, \textit{Ja-Be-Ja}  \cite{rahimian2013ja} (a local search partitioner), \textit{Fennel} \cite{tsourakakis2014fennel} (a streaming balanced partitioner), and \textit{Spinner} \cite{martella2017spinner} (a LP-based partitioner, which is currently deemed the state-of-the-art) are three \textit{vertex-centric} balanced partitioners. 

\section{Background} \label{Background}
    \subsection{Label propagation-based graph partitioning}
    \label{LPsection}
    
    LP \cite{raghavan2007near} is an iterative semi-supervised machine learning algorithm that infers unlabeled data from available labeled data. It has been used for community detection \cite{barber2009detecting} and balanced graph partitioning \cite{martella2017spinner}. Spinner \cite{martella2017spinner} uses LP to solve $k$-way graph partitioning via producing $k$ scores for $k$ partitions for every vertex. Afterwards, it chooses for each vertex the partition with the maximum score as a candidate partition. Each vertex will then migrate to its candidate partition only if the specified balance across partitions is not impacted. To elaborate, the scoring function of Spinner \cite{martella2017spinner} is as follows:
    %\abovedisplayshortskip=-5pt
    %\belowdisplayshortskip=-5pt
    %\setlength\abovedisplayskip{-1pt}    
    %\setlength\belowdisplayskip{2pt}    
    \begin{align}
        \widehat{score}(v,l) &= \sum_{u \in N(v)} {\frac{\widehat{w}(u,v) \cdot \delta(\widehat{\psi}(u),l))}{\sum_{u \in N(v)} \widehat{w}(u,v)} - \widehat{\pi}(l)} \label{eq:lp} 
        \\
        \widehat{w}(u, v) &=
        \begin{cases}
            1 \ \ \text{if $(u, v) \in E \oplus (v,u) \in E$} \\
            2 \ \ \text{if  $(u, v) \in E \wedge (v,u) \in E$}
        \end{cases} 
        \label{eq:weight}
        \\
        \widehat{\pi}(l) &= \frac{b(l)}{C}, \ b(l) = \sum_{u \in B(l)}{deg(u) \cdot \delta(\widehat{\psi}(u), l)} \label{eq:pi}
    \end{align}
    
    where in (\ref{eq:lp}), an edge $e \in E$ is a pair $(u, v)$ with $u, v \in V$, $N(v) = \{u | u \in V, (u, v) \lor (v, u) \in E\}$ is the neighborhood of vertex $v$, $\delta$ is the Kronecker delta where $\delta(i,j) = 1$ if $i = j$ and 0 otherwise, $\widehat{\psi}$ is a labeling function where $\widehat{\psi}: V \rightarrow L$ such that $\widehat{\psi}(v) = l$; if $v \in B(l)$, $\widehat{w}(u,v)$ is the weighing function computed in (\ref{eq:weight}), $\widehat{\pi}(l)$ is a penalty function with $b(l)$ being the load of partition $l$, and capacity $C = (\epsilon \cdot |E|) / k$.
    
    For each vertex $v$ in Spinner, the partition with the maximum score is considered as $v$'s candidate partition (say, $l$) and $v$ may migrate to $l$ only if the probability of migration to $l$, $\widehat{p}(l)$,  is greater than a random number generated between $[0,1]$. More precisely, the probability of migration of $v$ to $l$, is calculated based on $l$'s remaining capacity, $r(l) = C - b(l)$, divided by the number of candidate edges in $l$, $m(l) = \sum_{u \in M(l)} deg(u)$, where $M(l)$ is the set of vertices to migrate to $l$ and $deg(u)$ is the degree of each vertex in this set.

    \subsection{Mathematical framework of learning automata}
    \label{LA}
    Learning automaton \cite{narendra2012learning} is a probabilistic decision-making algorithm belongs to the family of Reinforcement Learning (RL). It draws an action using its probability distribution and applies it to the environment. By taking a sequence of actions and receiving reactions, learning automaton learns an optimal action. The common type of the learning automaton is variable structure learning automata, which is defined using the quadruple $[A(n), P(n), R(n), T]$ where $n$ is the learning step and: 1) $A(n) \in \{a_1, ..., a_m\}$ is the set of actions with $m$ being the number of actions, 2) $P(n) = \{p_1(n), ..., p_m(n)\}$ such that $\sum_{i=1}^{m}{p_i = 1}$ is the probability vector where the action taken in step $n$ is chosen using a roulette wheel \cite{goldberg1990probability}, 3) $R(n) = \{r_1(n), ..., r_m(n)\}$ wehere $r_i \in \{0, 1\}$ is the reinforcement signal, and 4) $T$ is the linear learning algorithm where in $n^{th}$ step $P(n+1) = T[A(n), P(n), R(n)]$.

    In $n^{th}$ step of $T$, if action $a_i(n)$ receives reward signal $r_i(n) = 0$, probability vector $P(n)$ is updated as follows:
    \abovedisplayshortskip=-2pt
    \belowdisplayshortskip=-2pt
    \begin{align}
        p_j(n+1) &= 
            \begin{cases}
                    p_j(n) + \alpha (1 - p_j(n)) &\text{$j = i$}\\
                    p_j(n) (1 - \alpha) &\text{$j \neq i$}
            \end{cases}
            \label{eq:la_reward}
    \end{align}
    
    Otherwise, if action $a_i(n)$ receives penalty signal $r_i(n) = 1$, probability vector $P(n)$ is updated as follows:
    \abovedisplayshortskip=-2pt
    \belowdisplayshortskip=-2pt
    \begin{align}
        p_j(n+1) &= 
            \begin{cases}
                p_j(n) (1 - \beta) &\text{$j = i$}\\
                p_j(n) (1 - \beta) + \frac{\beta}{m - 1} &\text{$j \neq i$}
            \end{cases} 
            \label{eq:la_penalty}
    \end{align}

    where in (\ref{eq:la_reward}) and (\ref{eq:la_penalty}), $\alpha$ and $\beta$ are reward and penalty parameters, respectively.

\section{Revolver} \label{revolver}
    Revolver is an application of Reinforcement Learning (RL) to graph partitioning \cite{mofrad2018revolver}. It uses weighted learning automata (Section \ref{wla}) to partition a graph. In Revolver, a new normalized Label Propagation (LP) algorithm (Section \ref{fullnlp}) is extrapolated to form an objective function which produces weights that express the decency of the assigned partitions by Learning Automata (LA). Moreover, Revolver partitions the graph in a vertex-centric manner. In particular, each vertex {\em pulls} information from its neighboring vertices to calculate a score for each partition, before {\em pushing} the calculated scores (as weights) back to them. Subsequently, reinforcement signals are computed at each vertex based on the accumulated weights gathered from all the vertex's neighbors. Finally, the probabilities of actions associated with partitions are updated using weights and reinforcement signals accordingly.
    
    In the following sub-sections, we introduce our extension of LA for graph partitioning, namely, weighted LA. Afterwards, we introduce our new normalized LP formulas, before showing how graph partitioning can be solved using LA. Lastly, we discuss how LP is used to train LA for graph partitioning. 
    
\subsection{Weighted Learning Automata} \label{wla}
    A typical learning automaton uses (\ref{eq:la_reward}) or (\ref{eq:la_penalty}) for updating its probability vector. However, in a complex environment such as multimillion-node graphs with a large number of actions $m$ (or partitions in a partitioning problem) this updating strategy tends to fail because of two reasons: 1) at any given step, (\ref{eq:la_reward}) or (\ref{eq:la_penalty}) can reinforce only one of the actions, meaning in conventional learning automaton there is only one reward signal and the rest are penalty signals,  and 2) given that the initial probability vector is initialized with $1 / m$, for a large number of actions this value will converge to zero (i.e., $\lim_{m\to\infty} 1 / m = 0$), which means a considerable amount of time will be required to reach a consensus on a single action.
    
    The above limitations motivate us to propose the \textit{Weighted Learning Automata}, which is able to distribute the reinforcement signals among an entire set of actions rather than concentrating it on a single action. Consequently, a weight vector $W$ is added to support the weighted probability updates for reward and penalty signals. Therefore, while updating the $i^{th}$ element of the probability vector using its weight and reinforcement signal, the rest of elements should also be updated using their weights and the negation of the $i^{th}$ reinforcement signal.
    
    Thus, the weighted learning automaton can be defined using the quintuple $[A(n), P(n), R(n), W(n), T]$, where $n$ is the learning step and: $A(n)$, $P(n)$, and $R(n)$ are the same as before (see Section \ref{LA}), $W(n) = \{w_1(n), ..., w_m(n)\}$ is the set of weights for reward and penalty signals, $w_i \in [0, 1]$ and $\sum_{i = 1}^{m}{w_i = 2}$ (more on this shortly), and $T$ is the linear learning algorithm, where in $n^{th}$ step $P(n+1) = T[A(n), P(n), R(n), W(n)]$. 
    
    In $n^{th}$ step of $T$, if action $a_i(n)$ receives reward signal $r_i(n) = 0$, probability vector $P(n)$ is updated as follows:
    
    \begin{align}
        p_j(n+1) &=
            \begin{cases}
                p_j(n) + \alpha \cdot w_j(n) (1 - p_j(n)) &\text{$ j = i$}\\
                p_j(n) (1 - (\alpha \cdot w_j(n))) &\text{$j \neq i$}
            \end{cases}
            \label{eq:wla_reward}
    \end{align}
    
    Otherwise, if action $a_i(n)$ receives penalty signal $r_i(n) = 1$, probability vector $P(n)$ is updated as follows:
    \abovedisplayshortskip=-2pt
    \begin{align}
        p_j(n+1) &=
                \begin{cases}
                    p_j(n)(1-(\beta \cdot w_j(n) )) &\text{$j = i$} \\
                    p_j(n) (1 - (\beta \cdot w_j(n))) + \frac{\beta}{m - 1} &\text{$j \neq i$} 
        \end{cases}
        \label{eq:wla_penalty}           
    \end{align}
    
    where $\alpha$ and $\beta$ are the reward and penalty parameters, respectively and $W(n)$ contains weights for reward and penalty reinforcement signals $R(n)$. To guarantee the correctness of the calculations, the sum of weights for reward signals and penalty signals should be both 1 (i.e., $\sum_{i = 1}^{m}{w_i \delta(0, r_i) = 1}$ and $\sum_{i = 1}^{m}{w_i \delta(1, r_i) = 1}$ where $\delta$ is Kronecker delta), subsequently, $\sum_{i = 1}^{m}{w_i = 2}$. As such, compared to the original learning automaton where the probability is updated $m$ times via multiple passes of (\ref{eq:la_reward}) or (\ref{eq:la_penalty}) (having only one reward reinforcement signal), in weighted learning automaton, (\ref{eq:wla_reward}) or (\ref{eq:wla_penalty}) are executed $m ^2$ times in total ($m$ times for each $r_i$) so as to apply $m$ different reward or penalty reinforcement signals and keep the sum of probabilities equals to 1.

\subsection{Normalized $k$-way label propagation for graph partitioning}  \label{fullnlp}
    In a multi-term LP, a dominant term easily causes huge variations in score computations. Normalization of terms is a solution for this. Thus, we propose a normalized LP, which consists of a normalized \textit{weighing term} $\tau$ and \textit{penalty term} $\pi$ defined as follows:
    \abovedisplayshortskip=-2pt
    \begin{align}
        score(v,l) &= \frac{\tau(v,l) + \pi(l)}{2} \label{eq:nlp2}
        \\
        \tau(v,l) &= \sum_{u \in N(v)} {\frac{\widehat{w}(u,v) \cdot \delta(\widehat{\psi}(u),l)}{\sum_{u \in N(v)} \widehat{w}(u,v)}} \label{eq:nlp1}
        \\
        \pi(l) &= \frac{1 - (b(l) / C)}{\sum_{i = 1}^{k}{1 - (b(l_i)/C)}} \label{eq:pin}
   \end{align}
   
    where in (\ref{eq:nlp2}), a score is produced for the $l^{th}$ partition of vertex $v$, $\tau$ is normalized based on the total weight of $v$'s neighborhood $N(v)$, $\pi$ (which produces penalties for partitions) is normalized based on the total load of the system\footnote{Note, if there is a negative penalty, penalties will be augmented with respect to the minimum negative value before normalization.}, $\delta$ is Kronecker delta, and $\widehat{w}$ and $\widehat{\psi}$ are defined in Section \ref{LPsection}.
 
    \subsection{How to use reinforcement learning for graph partitioning}
\begin{figure}[t]
        \begin{center}
        \includegraphics[width=\linewidth,height=0.27\linewidth,keepaspectratio]{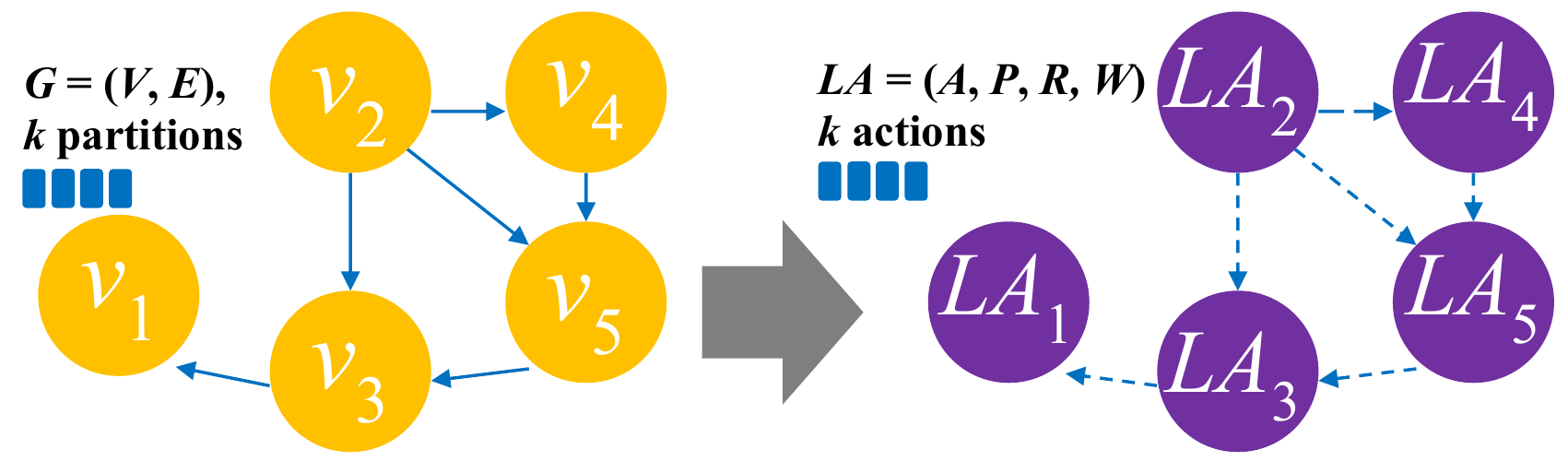}
        \end{center}
        \squeezeup
        \caption{LA with $k$ actions are allocated for nodes of graph. The rectangles represent LA actions associated with partitions.
        }
        \squeezeup
        \label{fig:GtoLA}
    \end{figure}

    To partition a graph $G = (V, E)$ into $k$ disjoint partitions, $LA=[A, P, R, W)]$ is utilized, where a learning automaton is associated with each $v \in V$ and the available range of partitions constitutes the action set of learning automaton. The mapping is shown in Figure \ref{fig:GtoLA} where a network of LA is created from a hypothetical graph $G$.
    
    The way LA is laid-out to solve a partitioning problem goes as follows: 1) a network of LA is created where $|V| = |LA|$ with one LA per each vertex in $V$, 2) a learning automaton can find neighboring LA using a subset of $E$, which belongs to $v$, 3) in each step, the network of LA determines the partitions for vertices in parallel; the action set of a learning automaton is the same as the range of available partitions and the probability of actions $P$ is initialized to $1/k$, 4) scores that are generated from multiple passes of (\ref{eq:nlp2}) are evaluated by (\ref{eq:obj}) to form the weight vector $W$, 5) the reinforcement signal $R$ is, subsequently, constructed from the weight vector values and is used to measure the merit of the current partitioning configuration, alongside giving LA feedback via updating probabilities using (\ref{eq:wla_reward}) and (\ref{eq:wla_penalty}), and finally 6) LA will learn how to partition the graph by taking a series of actions and receiving reinforcement signals.

\subsection{How to use label propagation to train a reinforcement learning algorithm}  
    \subsubsection{LA action selection} Each step of Revolver starts with LA taking actions for determining the partitions of vertices locally (\circled{1} in Figure \ref{fig:scenario}). Each vertex has an analogous learning automaton in the network of LA with an action set equal to the available partitions. The LA determines the candidate partition for the vertex using a \textit{roulette wheel} populated by its probability vector. Actions with larger probabilities will have a higher chance for being selected by the automaton.

    \subsubsection{Calculating vertex migration probability} After taking actions by LA and selecting candidate partitions for vertices, the remaining load of partition $l$ is calculated using the edges currently assigned to it (i.e., $C - b(l)$) and the demanding load is computed using the number of candidate edges in $l$ (i.e., $\sum_{u \in M(l)} deg(u)$) (see Section~\ref{LPsection}). Lastly, to calculate the probability for a vertex to migrate to a candidate partition, we simply divide the remaining load by the demanding load.

    \subsubsection{Computing vertex score} 
\begin{figure*}
        \begin{center}
        \includegraphics[width=\linewidth,height=.3\linewidth,keepaspectratio]{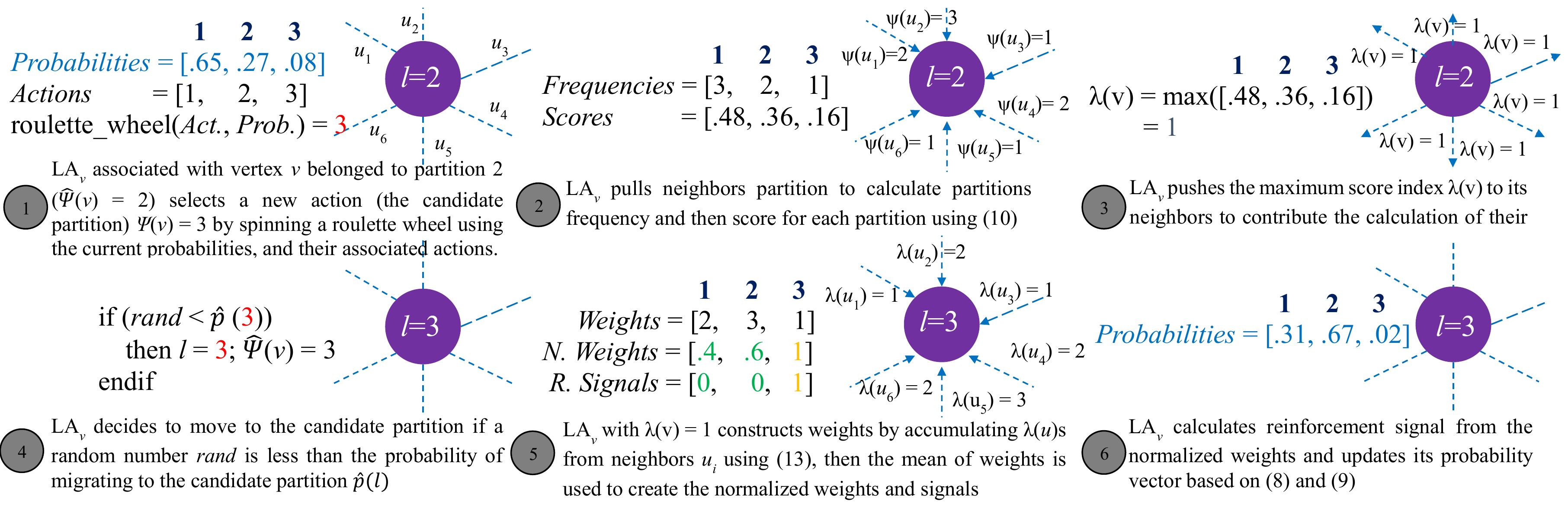}
        \end{center}
        \squeezeup
        \caption{An example of a 3-action learning automaton, $\text{LA}_v$, corresponding to vertex $v$, interacting with its neighbors through push and pull mechanisms to partition a graph into 3 sub-partitions. In the beginning of the step, the current label for $v$ is $l = 2$ and the probability vector for $\text{LA}_v$ is $[.65, .27, .08]$. At the end of the step, node $v$ is assigned to label 3 ($l = 3)$ and the probability vector of $\text{LA}_v$ is updated and changed to $[.31, .67, .2]$}
        \label{fig:scenario}
        \squeezeup
    \end{figure*}

The normalized LP in (\ref{eq:nlp2}) is used to calculate a score for each partition of a vertex based on the vertex's neighboring vertices partitions and the current load of the partition (\circled{2} in Figure \ref{fig:scenario}). To the contrary of other vertex-centric graph partitioning algorithms like Spinner \cite{martella2017spinner}, which migrates a vertex $v$ to the partition with the maximum score, Revolver utilizes a $\lambda$ function to extracts the index of the maximum score for future training of LA i.e. $\lambda(v) = \max_{l \in \{1, ..., k\}}{(score(v, l))}$. (\circled{3} in Figure \ref{fig:scenario}).
    
    \subsubsection{Executing vertex migration} The decision for migrating a vertex to a new partition is instrumented by comparing the selected action versus the current partition. If the two are not similar, a random number is generated against the migration probability of the candidate partition to determine whether to move the vertex to the new partition or not (\circled{4} in Figure \ref{fig:scenario}).

    \subsubsection{Evaluating the objective function} Vertex $v$ receives the maximum score label of its neighbors $\lambda(u) \ | \ \forall u \in N(v)$ and its learning automaton updates its corresponding weight vector as follows (\circled{5} in Figure \ref{fig:scenario}):

    \begin{align}
       w(v, \lambda(v)) \pluseq
       \begin{cases}
            \widehat{w}(u,v) \ \ \ \text{if $\delta(\psi(v), \lambda(u)) = 1$}
            \\
            1 \ \ \ \ \ \ \ \ \ \ \  \text{else if $\widehat{p}(\lambda(v)) \ > 0 $}
       \end{cases}
       \label{eq:obj}
    \end{align}
    
    where in (\ref{eq:obj}), $w(v, l)$ is the $l^{th}$ element of the weight vector $W$ belogned to vertex $v$, $\psi$ extracts the partition label assigned to vertex $v$ by its learning automaton (i.e., $\psi: A \rightarrow L$ such that $\psi(v) = l$ if LA$(v)$ assigns $l$ to vertex $v$), $\delta$ is the Kronecker delta, and $\widehat{w}$ and $\widehat{p}$ are defined in Section \ref{LPsection}.
    
    In (\ref{eq:obj}), the learning automaton associated with $u \in N(v)$ will receive a reinforcement signal proportional to $w(u,v)$ in two cases: 1) if the selected action of $\text{LA}_v$ is equal to $\lambda(v)$, or 2) if the migration probability of the selected action (or partition) $l$ is positive. Hence, these two cases try to reinforce actions associated with the highest score while preserving the balance by taking into account the probability of migration.

    \subsubsection{Constructing the reinforcement signals}
    The weight vector $W$ is a vector of weights populated by vertices belonged to $N(v)$. This vector shows the decency of partitions in $N(v)$, where higher weights represent the more promising partitions. To differentiate between favorable and unfavorable partitions while constructing the reinforcement signal $R$, we divide $W$ into two parts using its mean. Specifically, if $w_i$ is larger than the mean of weights, $r_i = 0$ (reward signal); otherwise $r_i = 1$ (penalty signal) (\circled{5} in Figure \ref{fig:scenario}).
    
    Lastly, each half of $W$ needs to be normalized independently so as the sum of each half becomes 1 and the sum of weight vector $W$ becomes 2. Normalized weights of reward and penalty reinforcement signals are necessary to keep the sum of LA probabilities equal to 1 while using (\ref{eq:wla_reward}) and (\ref{eq:wla_penalty}).
    
    \subsubsection{Updating learning automata probability vector}    
    After calculating weights and signals, the probability vector of a vertex $v$ is updated using (\ref{eq:wla_reward}) and (\ref{eq:wla_penalty}) (\circled{6} in Figure \ref{fig:scenario}). 
    \subsubsection{Updating remaining capacity}
    Calculating the remaining capacity is a simple subtraction of partition capacity $C$ from the current load of partition $b(l)$ at the end of each step. 
    
    \subsubsection{Checking convergence}
    Finally, Revolver halts if for a specified number of consecutive steps, score has not improved (i.e., $(S^i - S^{i-1}) < \theta$, where $\theta$ is the min score difference.)

\section{Experiments} \label{experimens}
\subsection{Experimental Environment}
    We conducted our experiments on a cluster of 96 nodes (especially that we implemented two versions of Revolver, distributed and parallel ones as noted shortly), each with 64 GB RAM and a 28 cores Intel Xeon CPU E5-2690 running at 2.60 GHz (Broadwell). The operating system of each node is Red Hat Enterprise Linux Server 7.3 (Maipo) with Linux kernel 3.10.0. Communication between nodes is achieved using Intel (R) Omni-Path with channel speed of 100 GB/s.
    
\subsection{Datasets}  
    Table \ref{table:dataset} reports the selected graphs along with their numbers of vertices and edges, densities, and skewnesses. In Table \ref{table:dataset}, the density of a graph $G(V, E)$ is calculated using $D= |E| / (|V| \cdot (|V|-1))$, and Pearson's 1\textsuperscript{st} skewness coefficient is computed using $(\mu - m) / \sigma$, where $\mu, m,$ and $\sigma$ are the mean, mode, and standard deviation of the outdegree edges. Negative or positive values of density illustrates to what degree a graph is skewed (i.e., whether toward left or right of the outdegree edges). From nine graphs, SO \cite{snapnets} and EU \cite{BoVWFI} are almost skew-free, USA \cite{diamacs} is uniquely left-skewed, and the rest are right-skewed, whereby they follow a power law distribution. Note that we use different graphs with varied degrees of skewness so as to comprehensively demonstrate the performance of Revolver. 
    
    %\hl{Finally, these graphs encompasse different degrees of skewness, so they can comprehensively represent performance characteristics of Revolver.}

    \begin{table}
    \begin{center}
        \caption{Graph datasets with numbers of \textbf{V}ertices, \textbf{E}dges, \textbf{D}ensities ($\times 10^{-5}$), and Pearson's 1\textsuperscript{st} \textbf{s}kewness coefficients.}
        \label{table:dataset}
        \begin{tabular}{c c c c c c}
            \hline
            Graph & $|V|$ & $|E|$ & $D$ & Skew
            \\
            \hline
            \multicolumn{1}{l}{Wiki-topcats (WIKI) \cite{snapnets}} &  1.79M & 28.51M & 0.88 & +0.35
            \\
            \multicolumn{1}{l}{UK-2007@1M (UK) \cite{BoVWFI}} &  1.00M & 41.24M & 4.12 & +0.81
            \\
            \multicolumn{1}{l}{USA-road (USA) \cite{diamacs}} &  23.9M & 58.33M & 0.01 & -0.59
            \\
            \multicolumn{1}{l}{Stackoverflow (SO) \cite{snapnets}} &  2.60M & 63.49M & 0.93 & +0.08
            \\
            \multicolumn{1}{l}{LiveJournal (LJ) \cite{snapnets}} &  4.84M & 68.99M & 0.29 & +0.36
            \\
            \multicolumn{1}{l}{EN-wiki-2013 (EN) \cite{BoVWFI}} &  4.20M & 101.3M & 0.57 & +0.35
            \\
            \multicolumn{1}{l}{Orkut (OK) \cite{snapnets}} & 3.07M & 117.1M & 1.24 & +0.29
            \\
            \multicolumn{1}{l}{Hollywood (HLWD) \cite{BoVWFI}} & 2.18M & 228.9M & 4.81 & +0.32
            \\
            \multicolumn{1}{l}{EU-2015-host (EU) \cite{BoVWFI}} & 11.2M & 386.9M & 0.30 & +0.07
        \end{tabular}
        \squeezeup
    \end{center}
    \end{table}

\subsection{Implementation Details}      
    We implemented two versions of Revolver, a multi-threaded asynchronous one in C/C++ and a synchronous one in Giraph \cite{ching2015one}. To encourage reproducibility and extensibility, we made them both open-source\footnote{Revolver's code is available at: https://github.com/hmofrad/revolver}. However, we report the results of the asynchronous version for two reasons: 1) Revolver can benefit from incremental changes in partitions offered by the asynchronous computation model during the partitioning process, and 2) Revolver's C/C++ implementation efficiently balances the vertices among working threads via allocating each subset of vertices to a separate thread. In particular, the vertices, $V$, of every graph are divided into {\em chunks} of size $|V| / n$ (with $n$ being the number of threads) and each chunk is assigned a separate thread on a separate core.

\subsection{Algorithms}      
    We compared Revolver against a number of partitioning algorithms : 1) \textit{Spinner} \cite{martella2017spinner}, a vertex-centric graph partitioning algorithm, which uses an LP scoring function to determine a suitable partition for every vertex $v$, 2) \textit{Hash partitioning}, where $v\mod k$ is used to hash a vertex with numerical id $v$ to its designated partition, and 3) \textit{Range partitioning}, where $(v \cdot k) / |V|$ is utilized to map a vertex with id $v$ to its partition.

\subsection{Performance Metrics}
    To demonstrate the quality of partitioning, we borrowed two metrics from \cite{martella2017spinner}, namely: 1) $Local \ Edges = (\sum_{\forall (u, v) \in E}{\delta(\widehat{\psi}(u), \widehat{\psi}(v)))} / |E|$, which is the number of edges with both ends at the same partition divided by the total number of edges, and 2) $Max \ Normalized \ Load = Max \ Load / Expected \ Load$, where \textit{Max Load} is the number of edges assigned to the highest loaded partition $\max_{l \in \{1, ... ,k\}}(b(l))$ (see Section \ref{Problem}) and \textit{Expected Load} is $|E|/k$. Also, we note that $1 - Local \ Edges$ is equal to $Edge \ cuts = (\sum_{(u, v) \in E }{1 - \delta(\widehat{\psi}(u), \widehat{\psi}(v)})) / |E|$. Clearly, with a set of machines assuming a one-to-one mapping from nodes to partitions, these metrics illustrate the degree at which a given workload can uniformly harness the available computing resources and stress the communication medium at run time. To this end, we indicate that the execution time of the partitioning algorithm as a performance metric is less informative in this context since having a faster runtime does not necessarily lead to better partitions.

    %\hl{Having a set of machines with a one-to-one mapping from nodes to partitions, together these metrics show the degree in which an example workload uniformly harnesses the available computing resources and stresses the communication medium at runtime. Note that the execution time of the partitioning algorithm as a performance metric is not important as having a faster runtime does not certainly lead to better partitions.}

    Communication becomes a bottleneck if an algorithm requires to perform a massive amount of message passing after each execution step. To this end, the metrics \textit{local edges} and \textit{edge cuts} represent the amount of intra-partition and inter-partition interactions, respectively and assess the degree to which an application may require sending/receiving internal or external messages. Furthermore, for an iterative application, the runtime of a single step of execution is bounded by the computation done at the highest loaded machine, which adds up to the latency of the system as well. As such, the metric \textit{max normalized load} captures the extent to which the computation time is affected by the highest loaded partition. 
    
\subsection{Experimental Settings}
\begin{figure*}
        \begin{center}
        \includegraphics[width=\linewidth,height=.6\linewidth,keepaspectratio]{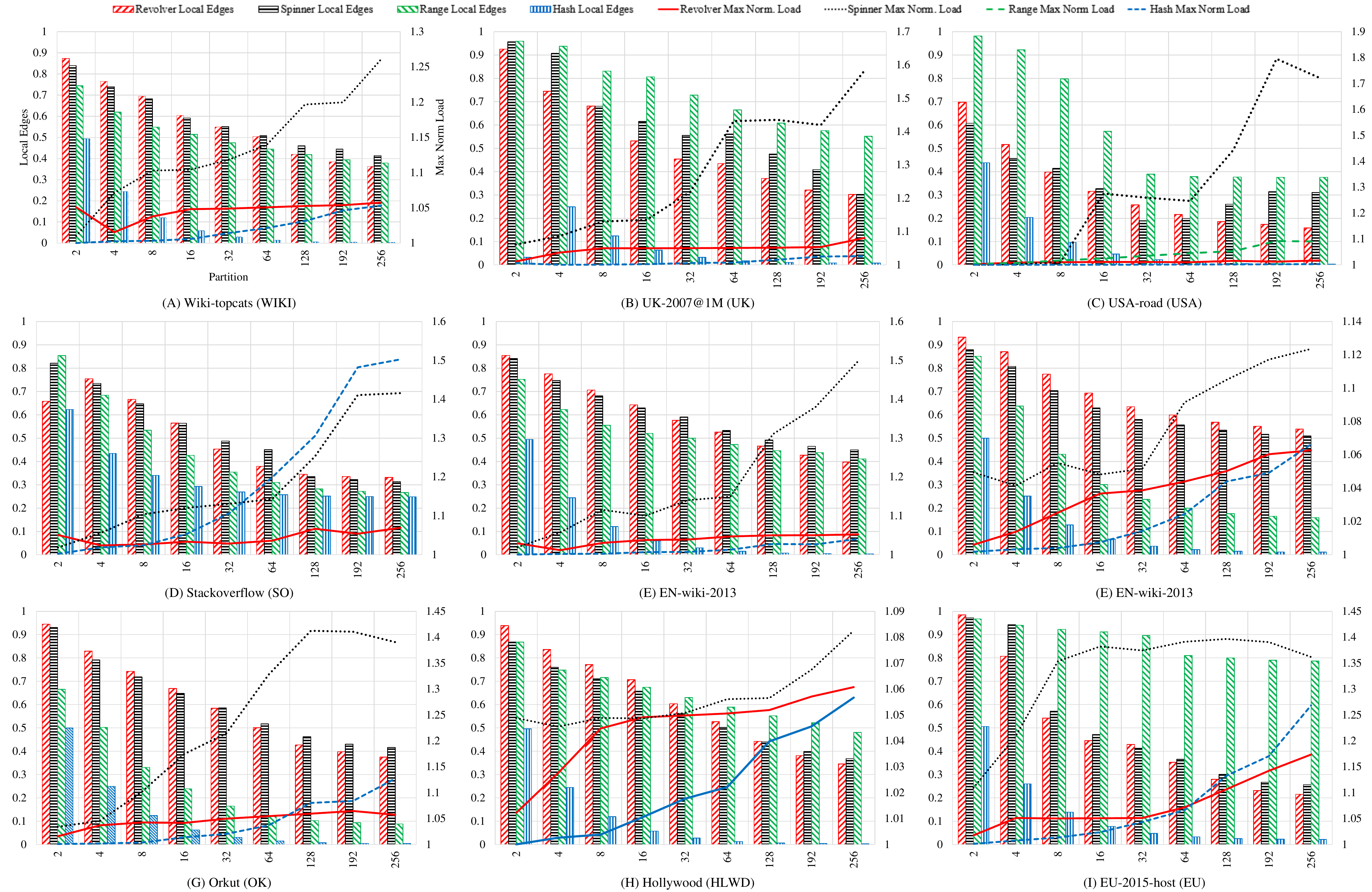}
        \end{center}
        \squeezeup
        \caption{Average local edges (bars scaled according to the left y axis) and max normalized load (lines scaled according to the right y axis) of Revolver, Spinner, Range, and Hash (Range max normalized load is removed intentionally from all charts except USA).}
        \squeezeup
        \label{fig:graphs}
    \end{figure*}

    In Figure \ref{fig:graphs}, we report the average \textit{local edges} and \textit{max normalized load} for 10 individual runs of each algorithm across different numbers of partitions 2, 4, 8, 16 32, 64, 128, 192, and 256. The results shown for Spinner are collected using Spinner's original implementation in Giraph \cite{martella2017spinner}. Moreover, for fair comparison, the same parameter settings as in \cite{martella2017spinner} are used to run Revolver and Spinner. Specifically, we set the max number of steps to 290, the max number of consecutive iterations for halting to 5, the min halting score difference to 0.001, and the imbalance ratio $\epsilon$ to 0.05. In addition, the LA reward and penalty parameters $\alpha$ and $\beta$ are set to 1 and 0.1, respectively. We note that the maximum normalized load for Range partitioning is so bad (e.g., Revolver has 60 times improvement compared to  Range with 256 partitions on EU graph); hence, we removed it from Figure \ref{fig:graphs} (except for USA graph) to avoid spoiling the plot scale.
    
\subsection{Analysis of Local Edges} \label{aole}
    We now discuss the results shown in Figure \ref{fig:graphs}. To start with, the clustered bars refer to local edges (left axis) and the lines denote max normalized load (right axis).
    
    \subsubsection{Right-skewed graphs} 
    WIKI, OK, LJ, EN, and HLWD are right-skewed graphs (see Table \ref{table:dataset}). Compared to Spinner, Hash, and Range, for WIKI (Figure \ref{fig:graphs}-A) and OK (Figure \ref{fig:graphs}-G), Revolver produces the best local edges with 2 - 64 partitions. Furthermore, it achieves the best local edges for LJ (Figure \ref{fig:graphs}-F), while maintaining 5\% improvement versus Spinner across all partitions. Alongside, it provides the best local edges for EN (Figure \ref{fig:graphs}-E) with 2 - 16 partitions. Lastly, compared to Spinner and Hash, Revolver produces the best local edges for HLWD (Figure \ref{fig:graphs}-H) with 2 - 128 partitions (almost 5\% improvement). In conclusion, Revolver adaptive strategy makes it a decent choice for partitioning right-skewed graphs, especially under smaller numbers of partitions.
    
    \subsubsection{Highly right-skewed graphs} UK is a highly right-skewed graph. Range produces the best local edges for UK, while Revolver accomplishes better max normalized load on this graph (Figure \ref{fig:graphs}-B). The right skewness feature of UK (Pearson's first rank = +0.81) indicates that the $mean$ of outdegree edges is far greater than of its $mode$, entailing that most of vertices have degrees less than the $mean$. Range partitions the graph based on the range of vertices and can exploit this UK's feature. Also, in comparing Revolver against Spinner, Revolver achieves better results with 256 partitions. 
    
    \subsubsection{Skew-free graphs}
    For skew-free graphs like SO and EU, Revolver produces the best local edges for SO (Figure \ref{fig:graphs}-D) with 4 - 16, 128 - 256 partitions, while Range provides the best local edges for EU (Figure \ref{fig:graphs}-I). Compared to EU, in SO, which is a denser graph with $D = 0.93 \times 10^{-5}$ (see Table \ref{table:dataset}), Revolver can produce better localized partitions for almost any number of partitions which shows it can effectively partition this type of graphs, independent of the number of partitions.

    \subsubsection{Left-skewed graphs} 
    For left-skewed graphs like USA (Figure \ref{fig:graphs}-C), Range produces better local edges and comparable max normalized load against Revolver. USA is a highly left-skewed graph (Pearson's first rank = -0.59) (see Table \ref{table:dataset}), which implies that outdegree edges are evenly distributed across vertices. Clearly, Range perfectly benefits from this characteristic and, accordingly, achieves superior results.
    
    \subsubsection{Impact of graph density and skewness on partitioning} To summarize the results of local edges, Revolver effectively partitioned both right-skewed and skew free graphs because its partitioning strategy is not highly depended to the way edges are distributed among vertices. Range exclusively partitioned highly right-skewed, dense skew free, and highly left-skewed graphs where edges are distributed evenly among vertices. For these kinds of graphs, Range's partitioning strategy simply extracts partitions from ranges of consecutive vertices while enjoying a balanced distribution of outdegree edges.

\subsection{Analysis of Max Normalized Load}  \label{aonl}
    \subsubsection{Trade-off between local edges and max normalized load}
    From Figure \ref{fig:graphs}-A all the way to Figure \ref{fig:graphs}-I (lines), Revolver always produces significantly better max normalized load compared to other algorithms, irrespective of the type of the graph. Unlike Spinner, Revolver's normalized LP does not allow the penalty function to vary the score independently and, subsequently, create unbalaced partitions. Revolver and Spinner leverage a 5\% imbalance ratio ($|E| / k \cdot ( 1 + \epsilon)$, where $\epsilon = 0.05 $), yet the largest partition produced by Spinner is always bigger than the allowed extra capacity. Evidently, this explains why Spinner accomplishes better local edges (i.e., because larger partitions will have more local edges). On the other hand, Hash produces comparable balanced partitions for these graphs with 2 - 64 partitions, while it always generates the worst local edges. Lastly, although Range provides the best local edges for UK, USA, and EU graphs, it achieves the worst max normalized load among all graphs (e.g., max normalized load of 1.6 - 60 times worst than Revolver for 2 - 256 partitions on EU), except for USA. Range is highly dependant on the way edges are distributed among vertices. The reason for why Range outperforms other algorithms on USA is that USA is a sparse ($D = 0.01 \times 10^{-5}$)  left-skewed graph with edges laid out evenly across vertices.
    
    \subsubsection{The impact of asynchronous processing}
    Since Revolver adopts an asynchronous computational model, the process of computing the scores of partitions and migrating a vertex to a new partition is executed \textit{on-the-fly}, whereby loads of the source and destination partitions are exchanged progressively. This relaxes the migration condition and enables Revolver to attain better max normalized load via utilizing the most recent changes in the partitioning configuration. Compared to Spinner, which is implemented synchronously, the asynchronous model of Revolver has a significant impact on the load distribution as shown in Figure \ref{fig:graphs}-A - Figure \ref{fig:graphs}-I (e.g., up to 28 $\times$ improvement in max normalized load on EU). 
    
\subsection{The Scalability Feature of Learning Automata}
    In a partitioning problem, as the number of partitions increases, the complexity of the problem grows as well. This is an inevitable outcome of the curse of dimensionality. In LA, as the number of actions increases, the initial probabilities are decreased which makes it harder to find optimal actions. Our weighted LA (see (\ref{eq:wla_reward}) and (\ref{eq:wla_penalty})) is designed to account for any increase in the dimensionality of the problem by having a weight for each element in the probability vector (which separates probability updates from the space complexity of the problem). Consequently, our weighted updating strategy guarantees a fair distribution of probabilities among the elements of the probability vector. This unique feature makes LA scalable and resistant to increases in the number of partitions.
    
\subsection{Convergence Characteristics of Revolver}
\begin{figure}
        \begin{center}
        \includegraphics[width=\linewidth]{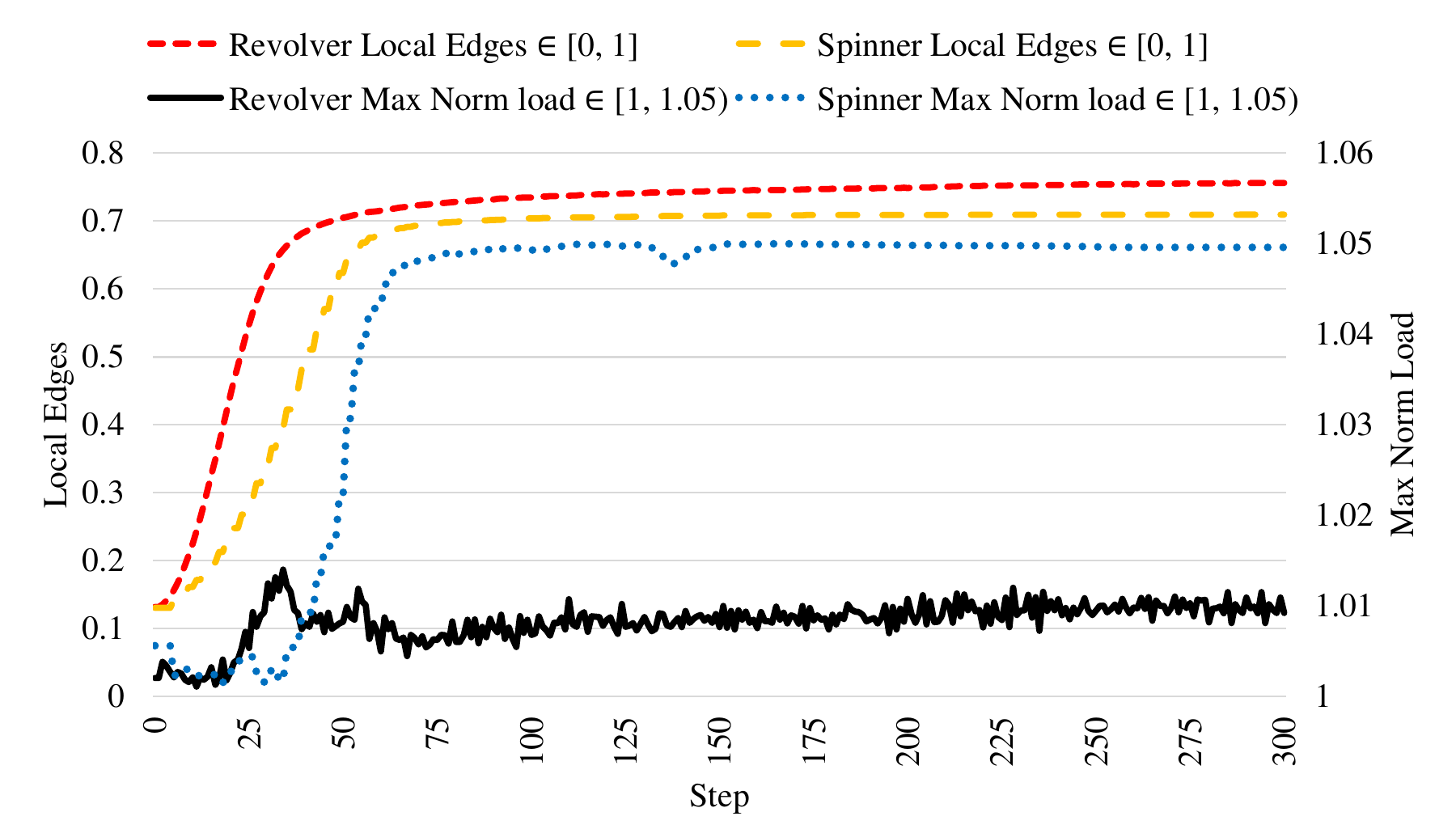}
        \end{center}
        \squeezeup
        \caption{Revolver convergence for LJ with 32 partitions across 290 steps.}
        \squeezeup
        \label{fig:conv}
    \end{figure}

    As Revolver and Spinner have clear advantages over Range and Hash in Sections \ref{aole} and \ref{aonl}, in Figure \ref{fig:conv}, we draw the convergence characteristics of local edges and max normalized load (left and right y axes) of Revolver and Spinner over LJ with 8 partitions and $\epsilon = 0.05$ (other graphs show a similar pattern to Figure \ref{fig:conv} and are not shown due to space limitations).
    
    Figure \ref{fig:conv} demonstrates an interesting observation about local edges. Specifically, Spinner's local edges become almost fixed after step 100, while Revolver keeps increasing local edges up to the end (step 300). This clearly indicates the strength of Revolver's adaptive strategy, which continuously allows reaching a consensus and does not get trapped in a local minimum. On the flip side, the greedy strategy of Spinner gets it trapped early on during execution. Moreover, there is a 5\% difference between the local edges produced by Revolver and Spinner, which further illustrates Revolver's superiority. In addition, Spinner stops increasing local edges when it fully utilizes the 5\% extra capacity ($\epsilon = 0.05$), while Revolver continues enhancing local edges, even without exhausting the entire available extra capacity.
    
    Comparing local edges and max normalized load, when Revolver hits the plateau of local edges after 30 steps, it harnesses up to 2\% extra capacity, whereas, Spinner local edges are only improved when more extra capacity is used (two middle lines of Figure \ref{fig:conv} shows this where local edges of Spinner is improved as a function of max normalized load).

    Figure \ref{fig:conv} also illuminates another pattern. In particular, Spinner tends to utilize its entire extra capacity in the first 75 steps, while Revolver barely consumes up to 2\% extra capacity during the whole run. The huge gap between Revolver's and Spinner's max normalized load is due to the fact that the asynchronous computation model of Revolver helps LA creating balanced partitions while utilizing significantly less extra capacity. In contrary, Spinner fails in achieving balanced partitions because of its strict synchronous model.

\section{Conclusions} \label{conclusions}
    In this work, we proposed Revolver, an asynchronous reinforcement learning algorithm capable of partitioning multimillion-node graphs. In Revolver, each vertex is assigned to an independent learning automaton to determine the corresponding suitable partition. In addition, a normalized label propagation algorithm is incorporated to asses partitioning results and provide feedback to learning automata. Experimental results show that Revolver can provide locally-preserving partitions, without sacrificing load balance.
    
\section{Acknowledgments}
    This publication was made possible by NPRP grant \#7-1330-2-483 from the Qatar National Research Fund (a member of Qatar Foundation). This research was supported in part by the University of Pittsburgh Center for Research Computing through the resources provided.

{
\bibliographystyle{IEEEtran}
\bibliography{draft}

}

\end{document}